\documentclass[
]{revtex4-2}

\usepackage{graphicx}
\usepackage{dcolumn}
\usepackage{bm}

\usepackage{newtxtext}
\usepackage{newtxmath}
\usepackage{natbib}
\usepackage{hyperref}
\usepackage{subcaption}
\usepackage{wasysym}

\begin{document}

\preprint{APS/123-QED}

\title{The seed-carrying stalk of the linden diaspore ensures autorotating flight}

\author{Annette Cazaubiel}
 \affiliation{University of Oslo, Department of Mathematics, Oslo 0851, Norway}

\author{Andreas Carlson}\email{Contact author: acarlson@uio.no}
 \affiliation{University of Oslo, Department of Mathematics, Oslo 0851, Norway}
\affiliation{%
Ume{\aa} University, Department of Medical Biochemistry and Biophysics, 901 87 Ume{\aa}, Sweden
}%

\date{\today}

\begin{abstract}
The dispersal of seeds by wind is one of the most evolved mechanisms plants use to invade new territories. Linden trees grow diaspores with a curved bract acting as a wing, where the seed pods are connected underneath by a stalk. Besides the seed-carrying capacity, the other functions of the stalk remain unknown. We demonstrate that the stalk of linden (genus {\em Tilia L.}) diaspores plays an essential role in their flight. The stalk length to wing span ratio is found to be the key parameter for facilitating steady autorotating flight in {\em Tilia x europaea} diaspores, effectively doubling their flight time compared to an unsteady tumbling motion. Flight experiments with biological and synthetic diaspores reveal a critical stalk length needed to induce autorotation, which correlates with analysis of collected biological samples and points to a possible evolutionarily selected trait. Flow measurements show that the vortical structures around the curved wing cause the elevated lift force during the autorotating flight.

\end{abstract}

\maketitle


\section{Introduction}
The flight organ of many tree diapores is a thin structure resembling a wing that generates an autorotating flight that facilitate wind driven dispersal \cite{Azuma1989}. One such prominent example is the diaspore grown by trees of the genus {\em{Tilia L.}}, commonly known as linden or lime trees, which can be found in most of the northern hemisphere. Linden trees have been of significant importance in streets and parks in Europe for centuries \cite{Bengtsson2005,Wolff2019,Pigott1992}. The most famous example is perhaps the alley \textit{Unter den Linden} in Berlin, whose name "under the lime trees" refers to the row of small-leaved linden trees forming the boulevard. Like other urban trees, they play an important role in the ecosystem and improve the air quality of cities \cite{Weber2013}. Adult trees can reach around $\sim$30 m above ground and it is essential for the diaspore to quickly start autorotating after detaching from the mother tree to maximise flight time. The placement of the fruit's mass relative to the flight organ is important to facilitate efficient flight \cite{Varshney2012}. A curious feature of dispersal units from Linden trees is that the main portion of their weight is the fruits found a distance below the bract, a specialised leaf, which functions as a wing, see Fig. 1. A peduncle, a stalk-like structure connects the fruits to the wing. We show that the length of this stalk might be a selected feature to ensure autorotation flight of the diaspore, which reduces descent speed and can help maximise flight time with increased likelihood to be carried away by the wind. 

Plants are affected by wind \cite{vogel1984drag, de2008effects, vogel2013comparative,vogel1989drag} and have evolved a wide range of geometric shapes of diaspores to help aid wind dispersal of their seeds \cite{burrows1975wind,Augspurger1986}, where these bear similarities to parachutes \cite{Cummins2018}, tumblers \cite{mccutchen1977}, single bladed auto-gyrators \cite{Azuma1989,Lentink2009}, gliders \cite{AZUMA1987}, and helicopters (whirling fruits) \cite{Rabault2019,Attenborough1995}. All of these strategies involve a diaspore equipped with a flight organ, which can maximise the flight time of the seed to allow for long travel distances, and as such increase the ecological fitness of the plants \cite{Nathan2002}. The autorotating motion of diaspores generates a large aerodynamic lift force, which leads to a significant reduction in descent velocity and consequently an increased flight time for the diaspore to be transported further by the wind. In the seminal work of Norberg \cite{Norberg1973} and Greene \cite{Greene1993}, a scaling prediction was presented for the descent speed of a falling fruit. By considering the mass flux through the area spanned by the rotating wing(s), they argued that a change in momentum must balance the weight of the fruit, which gives a descent speed proportional to the square root of the ratio between the fruit's weight and the area spanned by the rotating wing. However, details of the flow is essential to obtain a complete picture of the generated lift force. Investigations of samaras flight, single bladed autorotating fruits from the genus {\em Acer}, commonly known as maple, have identified a large leading-edge vortex (LEV) that shed from the rotating blade to be responsible for the unusually large lift force generated on the blade \cite{Lentink2009}. The rotating movement of samaras is caused by the torque due to the difference in the positions of the center of mass and the center of pressure, which is fine tuned \cite{Yasuda1997}. As the mass distribution changes, so does the flight mode, where autorotation can completely disappear \cite{Hou2024}. It should be noted that fossils from samaras were found to be double bladed, a design that does not induce autorotation and has vanished in today's single bladed samaras grown by different conifers \cite{stevenson2015}. The autorotation of samaras is not simply the movement of a windmill, but a complex coupling between the initial torque and the rigid body dynamic of the asymmetric wing \cite{Varshney2012}. If simplifying a wing to a flat body falling in a fluid, we know that the geometry of the body and the placement of its center of mass can induce a complex dynamics \cite{field1997chaotic}. It is a classical problem dating back to Maxwell \cite{Maxwell2011}, where the interaction between the falling object and its wake gives rise to a variety of flow regimes and instabilities \cite{Ern2012}. Cards, coins and disks falling in a fluid, are shown to transition from fluttering to tumbling, demonstrated in experiments and theory, and predicted by numerical simulations \cite{Mahadevan1999,ANDERSEN2005,Auguste2013,Tinklenberg_Guala_Coletti_2023,field1997chaotic}. 

Earlier work on helicopter seeds has shown that there is an optimal curvature to generate lift for autorotating fruits with curved wings \cite{Rabault2019,Fauli2019}. Some tropical trees can carry multiple seeds, which could be an evolutionary advantage \cite{Augspurger1983}. The curved wing geometry has been highlighted as one reason why the dispersal unit from the Phoenix tree is able to carry more than a single seed \cite{Gan2022}. Also the curved geometry of the linden bract allows it to carry multiple seeds, which may constitute an advantage for the reproduction of linden trees. It is the stalk (peduncle) that connects underneath the bract, which carries the fruits.

\section{Settling of linden diaspores}
\begin{figure*}
\begin{subfigure}{0.59\linewidth}
\includegraphics[width=0.99\linewidth]{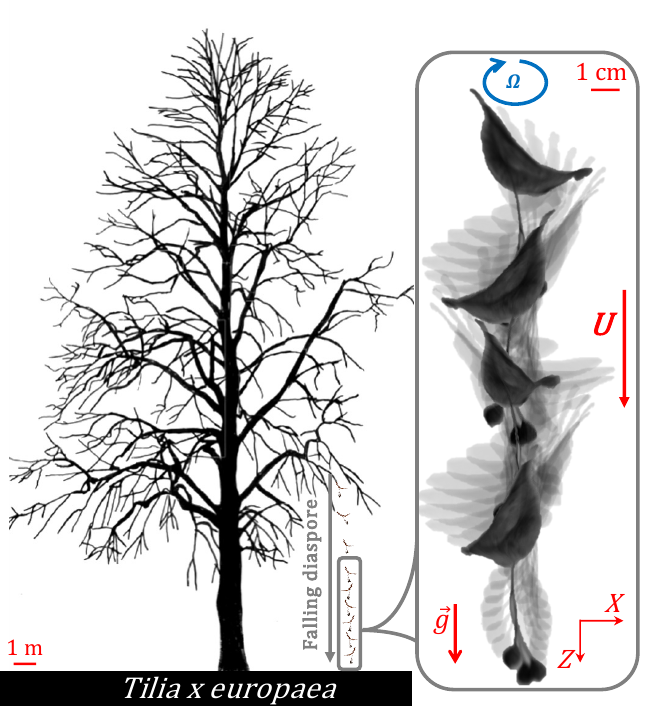}
\caption{}
\label{Fig1a}
\end{subfigure}
\begin{subfigure}{0.4\linewidth}
\includegraphics[width=0.99\linewidth]{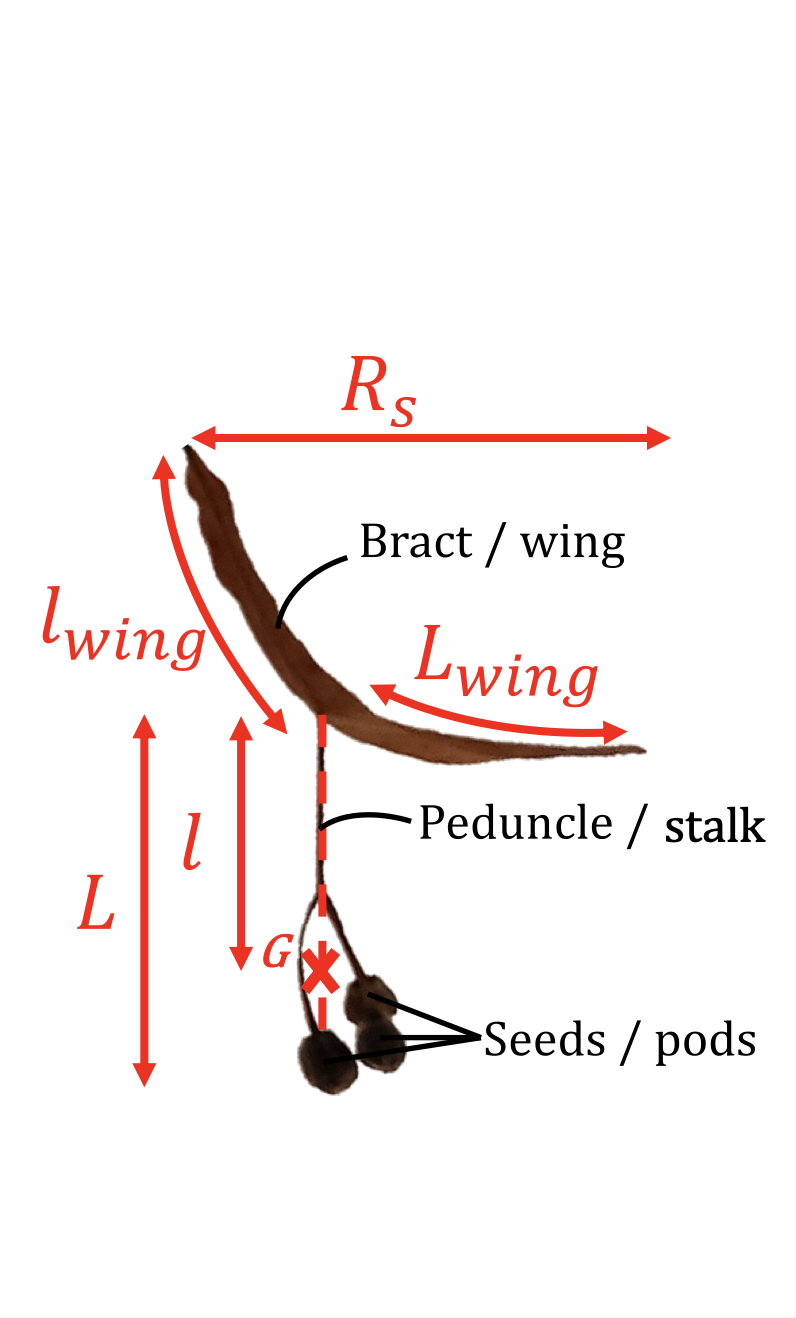}
\caption{}
\label{Fig1b}
\end{subfigure}
\centering
\begin{subfigure}{0.49\linewidth}
\includegraphics[width=0.99\linewidth]{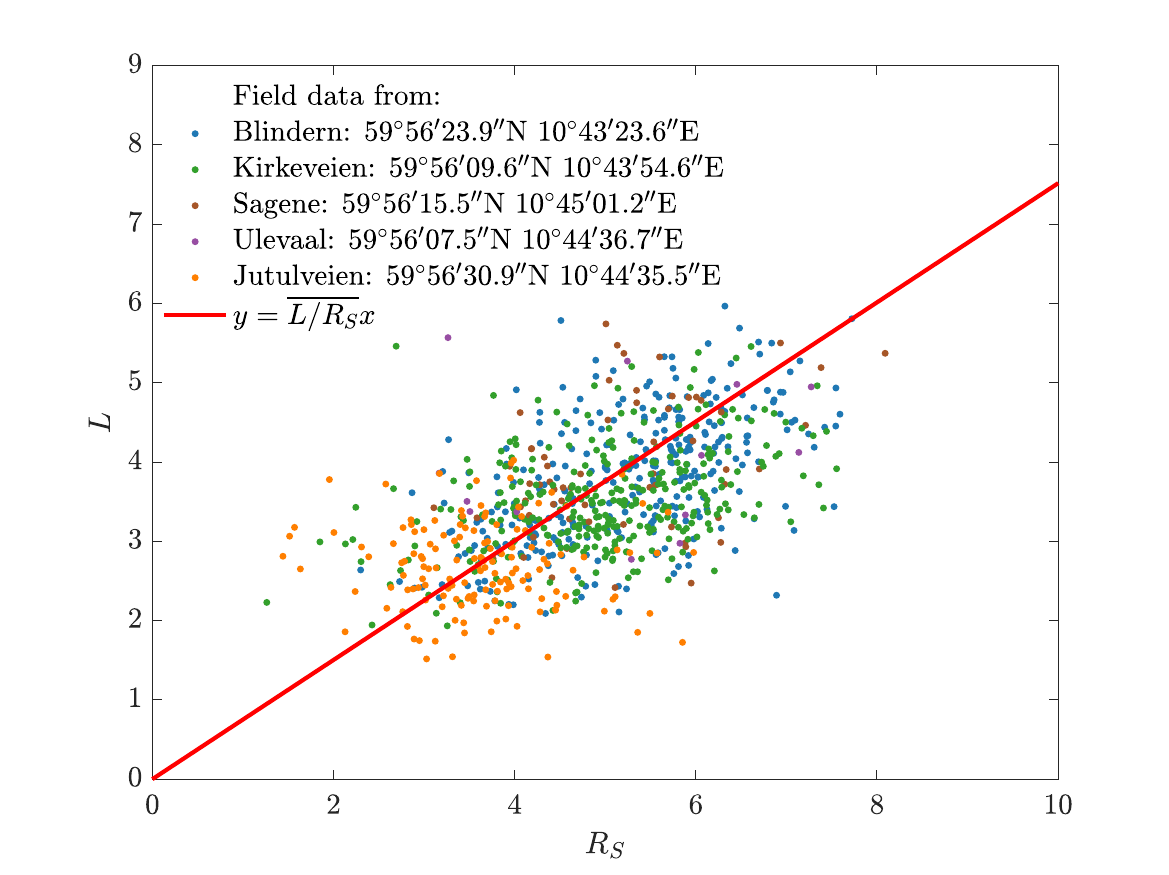}
\caption{}
\label{lRs}
\end{subfigure}
\begin{subfigure}{0.49\linewidth}
\includegraphics[width=0.99\linewidth]{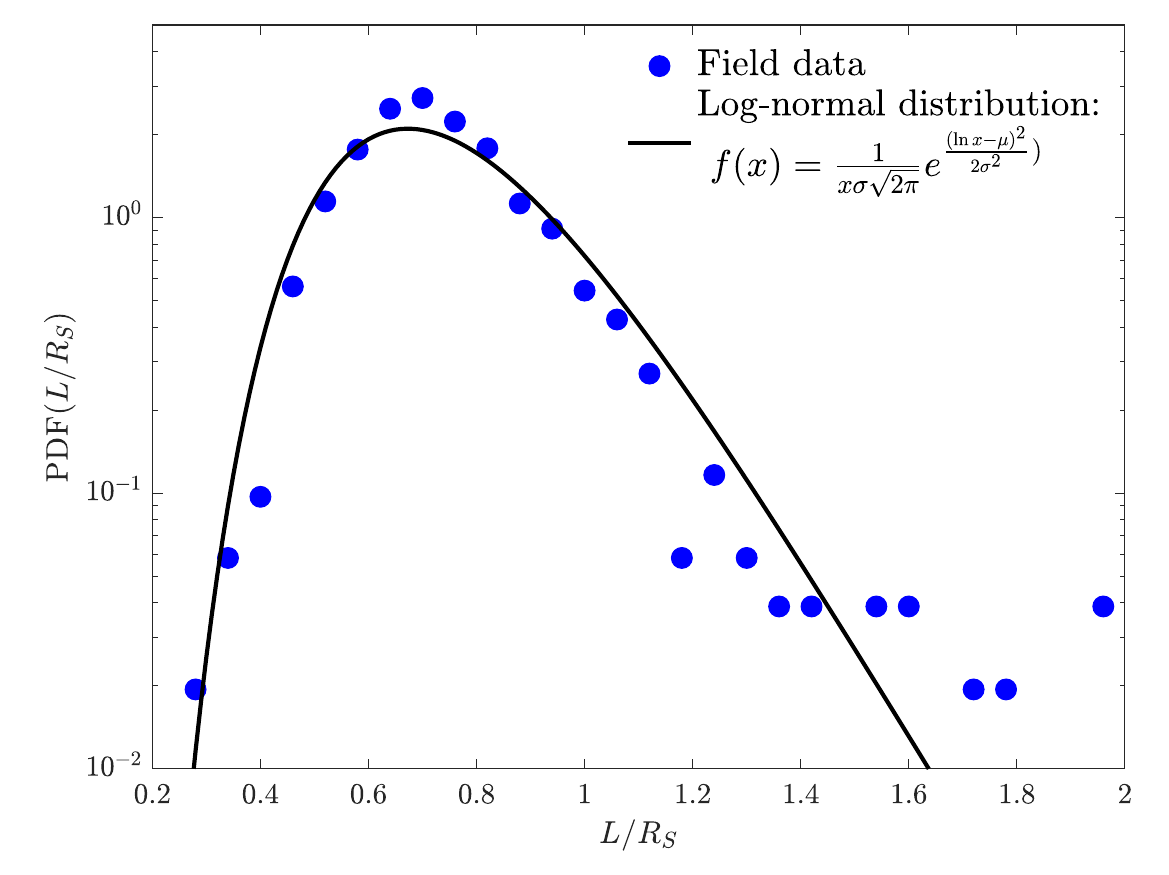}
\centering
\caption{}
\label{pdf}

\end{subfigure}
\caption{(a) Left: A sketch of an adult linden tree of the species \textit{Tilia x europaea} with a diaspore falling from its right side (not drawn on scale). Right panel: An illustration of the release of the autorotating diaspore through superposed pictures of a falling diaspore with a descent speed $U$, which rotates at angular velocity $\Omega$ around the vertical axis. (b) To characterize the geometry of the diaspore, we measure different characteristic lengths on the pictures of the collected biological samples. We determine the small and large lengths of the bract, respectively $l_{\text{wing}}$ and $L_{\text{wing}}$, the span $R_S$ of the wing, the total length of the stalk $L$ and the length between the bract and the center of mass of the diaspore $l$. (c) By collecting 859 biological samples from \textit{Tilia x europaea}, we obtain the statistical features of the geometrical properties of the diaspores. In particular, we show the length of the stalk, $L$ as a function of the wing span $R_S$. The solid red line represents the function $y=\overline{L/R_S}x$, with $\overline{L/R_S}$, the average of the ratio $L/R_S$. (d) The probability density function (PDF) is plotted for the ratio $L/R_S$ (blue dots). The solid black line represents the log-normal distribution:  $ f(x)= {1}/({x\sigma\sqrt{2\pi}})\exp({-\left({(\ln x -\mu)^2}/({2\sigma^2})\right)}),$
where $\mu=\ln(\bar{x}^2/\sqrt{\bar{x}^2+s^2})$ and $\sigma=\ln \left(1+\bar{x}^2/s^2\right)$, with $\bar{x}$=0.74 and $s$=0.18, the mean and the standard deviation of the distribution.}
\label{Fig1}
\end{figure*}

All trees from the genus \textit{Tilia L.}, called basswood, linden, or lime trees have diaspores with similar geometries, with solid round pods attached by a stalk to the center of a bract. Our study focuses on the most common species of linden tree, the \textit{Tilia x europaea} also known as common lime or common linden \cite{Hansen2020}. A mature tree typically reaches a height $H$, around 30 meters  \cite{Pigott1992} (see Figure \ref{Fig1a}). The crown has a conical, symmetrical shape, which starts about two to five meters above the ground \cite{Bengtsson2005}. When trees bloom, they form inflorescences, clusters of three to ten flowers attached together by a stalk about two to five centimeters long to a bract of about one centimeter wide and several centimeters long, see Fig \ref{Fig1}. The fruits are around half a centimeter in diameter and they detach from the tree together with the bract from early fall to very late in winter. The final geometrical properties of the bract and fruits, once released from the mother tree, are illustrated in Fig. \ref{Fig1b}. The bract is attached to the fruits by a stalk of length $L$ and $l$ is the distance between the wing and the center of mass of the diaspore. $m$ is the mass of the fruits and $M$, the total mass of the diaspore. The stalk is attached at the center of the bract in the chord direction and at the distance $l_\text{wing}$ from one end of the wing and $L_\text{wing}$ to the other end along the bract, see Fig. \ref{Fig1b}. $R_S$ is the span of the bract in the direction perpendicular to the stalk. When the diaspores settle, they quickly reach a steady autorotating state, with a nearly constant descent speed $U$ and a rotation speed $\Omega$ around a vertical axis, very close to the stalk (see Fig. \ref{Fig1a}).

We collected 859 diaspores of \textit{Tilia x europeae} from common linden trees in Oslo (more information on the protocol can be found in the Supplemental Materials (SM)). All samples have been imaged (Fig. \ref{Fig1b}) and these pictures have been used to measure the geometrical properties of the diaspore. The span $R_S$ ranges from 2 to 8 cm and the length of their stalk $L$ from 1.5 to 6.5 cm. We show in Fig. \ref{lRs}, $L$ as a function of $R_S$ for all the collected diaspores. There appears a moderate correlation (product moment correlation coefficient of approximately $0.55$) between the two lengths. Thus, there seems to be a link between the wing span and the length of the stalk. The probability density function shown in Fig. \ref{pdf} of the ratio $L/R_S$ is skewed to the right, following a log-normal distribution:
 $ f(x)= {1}/({x\sigma\sqrt{2\pi}})\exp({-\left({(\ln x -\mu)^2}/({2\sigma^2})\right)}),$
where $\mu=\ln(\bar{x}^2/\sqrt{\bar{x}^2+s^2})$ and $\sigma=\ln \left(1+\bar{x}^2/s^2\right)$, with $\bar{x}$=0.74 and $s$=0.18, mean and standard deviation of our distribution, respectively. Log-normal distributions have been found in many systems, ranging from biology to the social sciences \cite{koch1966logarithm,limpert2001log}. They stem from multiplicative variations, which can be the sign of evolution towards a more probable state \cite{gronholm2007natural}. This could indicate that linden diaspores have evolved towards a geometry, which is more probable.

To test the hypothesis that the ratio of stalk length to wing span $L/R_S$ is a key parameter to facilitate flight of diaspores from \textit{Tilia x europeae}, we perform flight experiments in air, with biological and synthetic diaspores. In these flight experiments, we systematically change the length of the stalk and the weight of the seed pods, while measuring the flight dynamics. This way, we control the vertical position of the center of mass and for the synthetic diaspores, we also control the shape of the wing allowing us to link the flight performance with the geometrical features of the diaspores. We observe that the length of the stalk sets the distance before autorotation starts for linden diaspores (see Fig. \ref{fig2a}). In order to examine how beneficial autorotation is for dispersal in the particular case of the curved wing of the linden diaspore, we measure the fluid flow around a 3D printed model diaspore \cite{Rabault2019, walker2021estimation} in water with a similar wing geometry using Particle Tracking Velocimetry or PTV (details found in the SM) to show the vortical structures around the synthetic wing. 
The Reynolds number, $Re=UR_S/\nu$, gives the ratio between the inertial and the viscous forces acting on the diaspore, and the Strouhal number, $St=R_sf/U$, is the ratio between the rotational and the descent speeds of the diaspores. $\nu$ is the kinematic viscosity of the ambient fluid and $f=\Omega/2\pi$. The biological samples of \textit{Tilia x europaea} have a descent speed $U\in [0.5,3]$ m/s and rotational speed $\Omega \in [600,1800]$ rpm, consistent with previous measurements \cite{Azuma1989} of \textit{Tilia miqueliana} Maxim. diaspores, i.e. $U=1.34$ m/s and $\Omega=832.6$ rpm. By using these numbers gives us $Re\approx3500-5000$ and $St\approx0.5-0.8$. We keep the Reynolds and the Strouhal numbers of the same order for our synthetic diaspores both in air and in 3D PTV measurements in water as further described in the SM, to ensure comparable flow structures.

\section{Results}

\begin{figure*}[t!]
\centering
\begin{subfigure}{0.47\linewidth}
\includegraphics[width=0.99\linewidth]{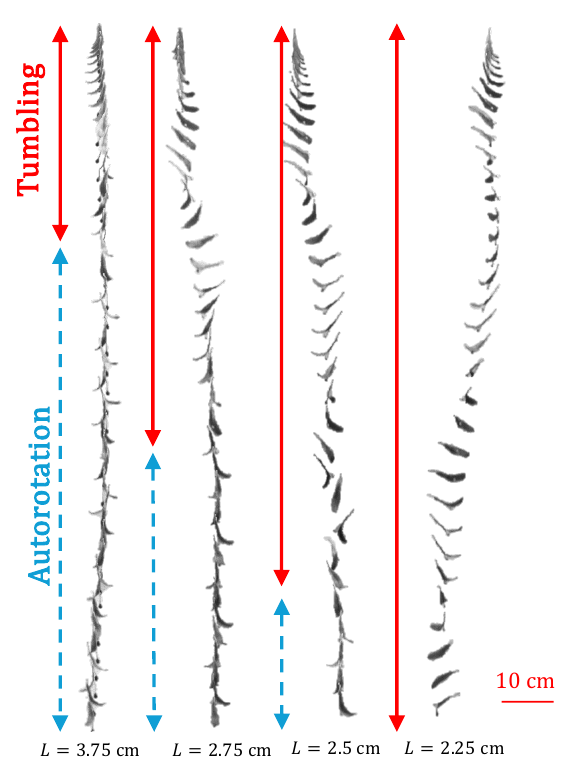}
\caption{}
\label{fig2a}
\end{subfigure}
\begin{subfigure}{0.52\linewidth}
\includegraphics[width=0.99\linewidth]{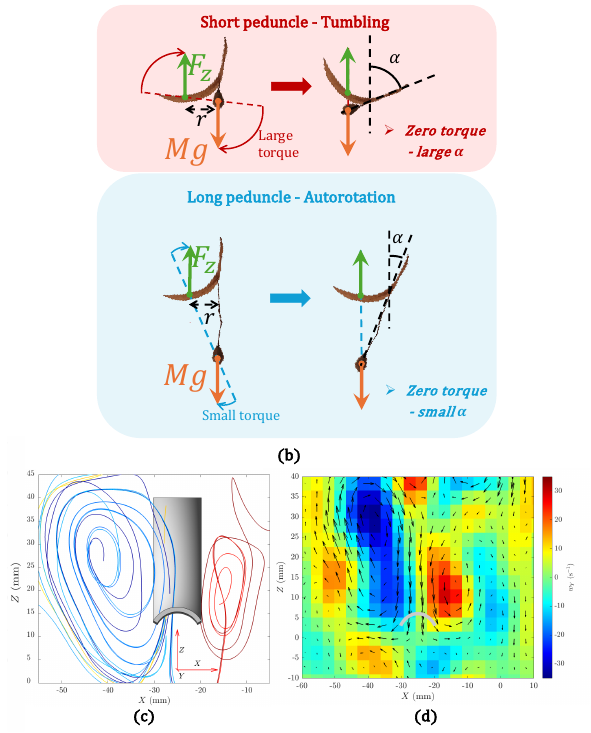}
\centering
\end{subfigure}
\caption{(a) Superposition of images of the flight of a diaspore from \textit{Tilia x europaraea}, where each image are separated by 20 ms. The panels from left to right show the effect of the length of the stalk with: $L=$[3.75 , 2.75, 2.5, 2.25] cm. The diaspore first accelerates and tumbles (illustrated by the solid red arrow), and then rotates around a vertical axis and decelerates (illustrated by the dashed blue arrow). The initial tumbling becomes longer when the stalk length $L$ decreases.  (b) Illustration of the torque acting on the diaspore as it tumbles for a short stalk (top) and a long stalk (on the bottom).  $F_z$ is the vertical component of the aerodynamic forces and $Mg$, the weight of the diaspore. (c) Streamlines of the 3D-flow around the wing of a synthetic diaspore falling in water seen from the front with $Re\approx4000$ and $St\approx0.65$. The wing is rotating from right to left during flight. A video of these streamlines around the rotating 3D wing is shown in the SM. The color map represents the vorticity, $w_Y$ in the $Y$ direction, with the same scale as in (d). The blue streamlines form the leading edge vortex and the red streamlines form the trailing edge vortex. (d), represents three vertical cuts of the velocity fields, at $Y=4$ mm (the tip of the wing is at $Y=0$ and the stalk at $Y=28$ mm). The colormap represents the vorticity in the $Y$ direction in s$^{-1}$. As shown also in the video (see SM) both by the velocity field and the vorticity map, there is two vortices attached to the wing, a strong leading edge vortex and a smaller trailing edge vortex. These vortices are concentrated towards the tip of the wing and leads to lift production.}
\label{fig2}
\end{figure*}

Flight experiments point to the importance of the wing-to-stalk length ratio for the flight of the diaspore. It appears that longer stalks are beneficial to initiate and stabilize autorotation of linden diaspores, as illustrated in Fig. \ref{fig2a} by superposed images taken every 20 ms of the flight of the same linden diaspore with different stalk lengths. Note that the vertical distance between two consecutive images of the falling diaspore is approximately twice as large during the tumbling state as during the autorotation, i.e., the descent velocity is reduced by half when the diaspores are autorotating. The velocity of the diaspore initially increases before it reaches a maximum and then decreases as autorotation sets in. 

For all stalk lengths, we can observe that, when released, the diaspore first accelerates and tumbles. For the largest stalk length (for the natural stalk length of the diaspore, $L=3.75$ cm), the diaspore quickly begins to rotate around a vertical axis after a short tumbling period and its descent speed decreases.
However, when we reduce the length of the stalk, both the fall distance before autorotation is initiated and the amplitude of the tumbling motion increase. The experiments reveal a quasi-periodic tumbling motion during which the diaspore does not decelerate. 
Eventually, for sufficiently short stalks, the diaspore tumbles for several meters and we do not observe any autorotation within the experimental window, i.e., the maximal descent distance of $9$ m. 

The altered flight behaviour going from steady-state autorotating to an unsteady tumbling as the stalk shortens can be explained by the torque in the direction tangential to the rotation. We already know that for a flat single bladed diaspore \cite{Varshney2012}, the difference in the horizontal positions of the center of mass and the center of aerodynamic pressure, $r$, imposes a torque on the wing, which causes the diaspore to tumble. For seeds attached to a flat wing, like for examples maple seeds, the geometry of the fruit and the weight distribution along the different axes render the tumbling motion unsteady. The energy of the tumbling motion then transfers to the spinning mode \cite{Varshney2012}, which generates a stable autoratation.  
Diaspores from the genus \textit{Tilia L.} have a stalk that plays the role of a lever arm and will attenuate the tumbling motion. As illustrated on Fig. \ref{fig2}b, for the same aerodynamic forces acting on the wing and the same weight of the fruits, the torque around the tumbling axis will be larger for a smaller stalk. Then as the diaspore starts tumbling, the torque will cancel out for a smaller tumbling angle for the diaspore with a long stalk than for the diaspore with a shorter stalk. This illustrates why the initial tumbling duration would be shorter for diaspores with longer stalk and why longer stalk would then be advantageous for facilitating a stable rotating motion around a vertical axis. To estimate the force and momentum balance of the diaspores, we decompose the bract into thin slices and calculate the aerodynamic forces on each slice assuming steady-state, using blade element theory \cite{Norberg1973,Rabault2019,Fauli2019}, see SM, which highlight that the angle of tumbling needs to be larger for shorter stalks.

\begin{figure*}

\begin{subfigure}{0.48\linewidth}
\includegraphics[width=.99\linewidth]{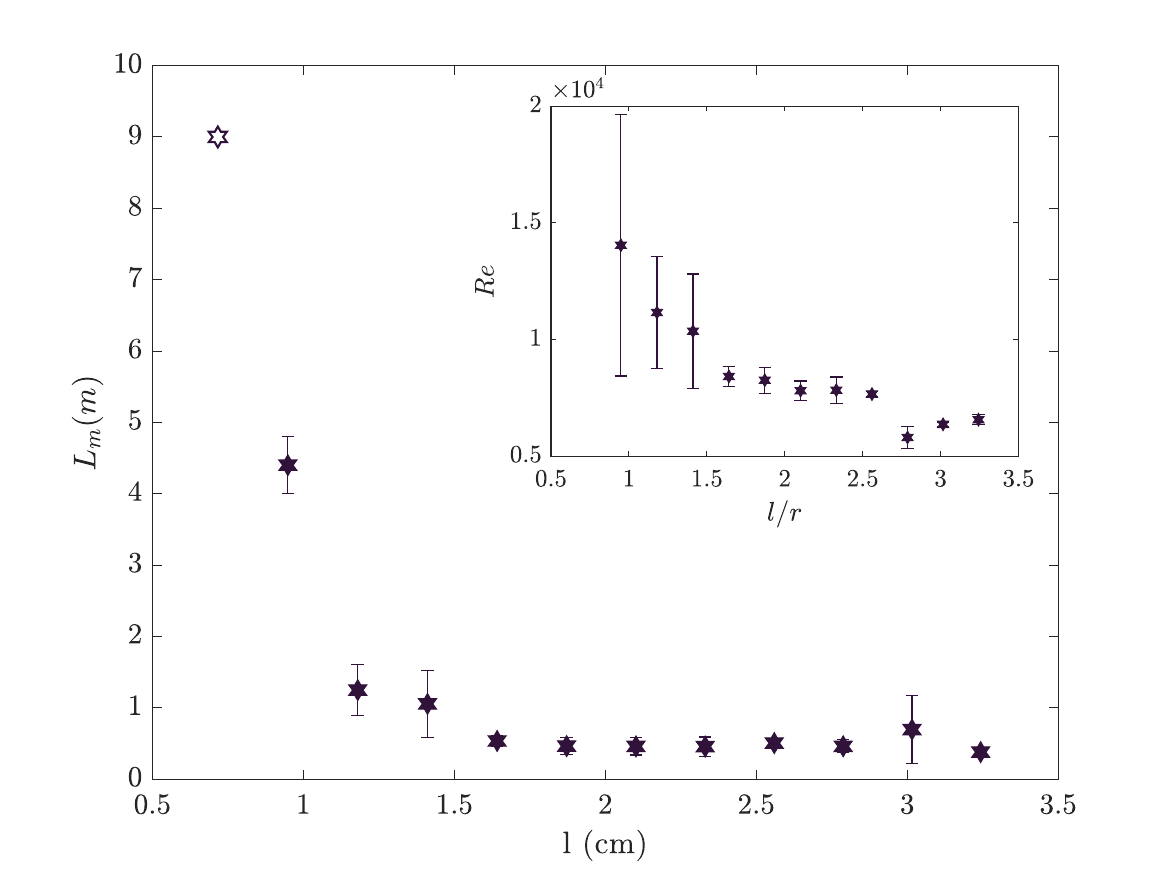}
\caption{}
\label{Fig3b}
\end{subfigure}
\centering
\begin{subfigure}{0.51\linewidth}
\includegraphics[width=0.99\linewidth]{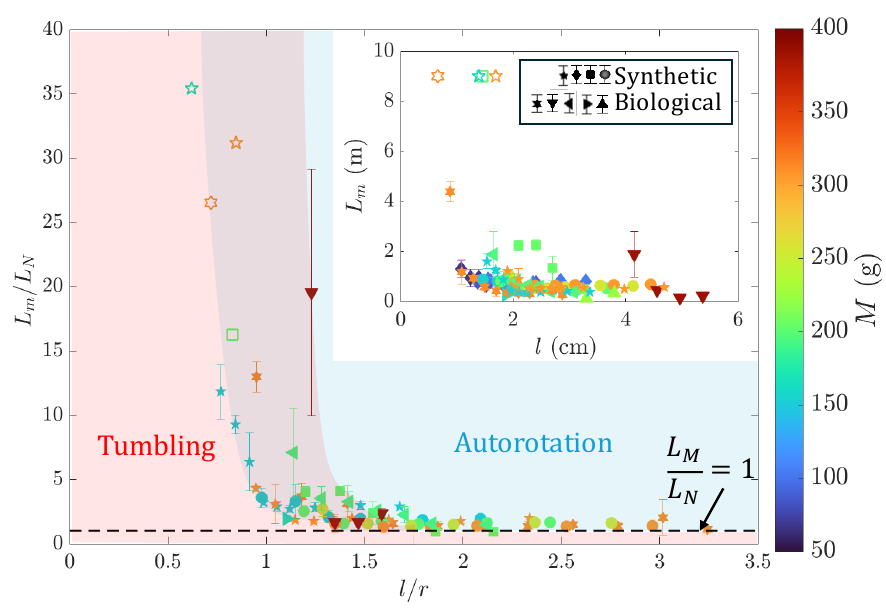}
\caption{}
\label{Fig3c}
\centering
\end{subfigure}

\caption{(a) $L_m$ as a function of the length of $l$, the hollow marker signifies that the autorotative steady state has not been reached at that length, inset: Reynolds number $R=UR_S/\nu$ as a function of $l$. (b) $L_m/L_N$ as a function of $l/r$, the color of the marker is set by the total weight of the diaspore as indicated by the colorbar. The different symbols signify different wing geometry, the different geometries are specified in a table in the SM. The hollow markers represent tumbling as no autorotation was measured in the experimental observation window and where $L_m$ then is set to be the decent height of the experiment ($9$ m). Inset: The dimensional data for $L_m$ is plotted as as a function of $l$.}
\end{figure*}

The aerodynamic forces applied on the wing also depend on the vortical structures of the flow around it. As shown by \cite{Lentink2009} a stable leading edge vortex would increase the lift force. We performed 3D measurements of the flow around a synthetic wing in water with $St\approx 0.65$ and $Re\approx 4000$ matched to the flight experiments in air. The instantaneous streamlines around the model wing are shown in 3D in Fig. \ref{fig2}c. Their color is scaled with the vorticity in the $Y$ direction.  We can observe a  leading edge vortex (the blue streamlines) and a smaller trailing edge vortex (the red streamlines) forming on the rotating wing. The velocity fields and the vorticity in the $Y$ direction in the $XZ$ plane are shown in Fig. \ref{fig2}d. They confirm the presence of the two vortices who will create an additional lift at the leading edge and at the trailing edge of the wing. Unlike maple seed \cite{Lentink2009}, those vortices are not concentrated near the center of rotation but near the tip of the wing, seemingly due to its curvature. The additional lift is thus also more concentrated towards the end of the bract increasing the torque applied on it and reinforcing the importance of a lever arm to balance it.

$L_m$ defines the descent distance, which is at the point of maximum velocity when autorotation is initiated. If there is no steady autorotation observed within the experimental window of $9$ m, we set $L_m=9$ m the height of the experiment and separate these point by using hollow markers in Fig. \ref{Fig3b} and Fig.\ref{Fig3c}. 
When we look at $L_m$ as a function of the vertical position of the center of mass of the diaspore $l$ (see Fig \ref{Fig3b}), we observe that $L_m$ decreases with the length of the stalk until it plateaus. It appears to diverge for small $l$, where five meters would approximately correspond to a fall height from the lower part of a crown of a linden tree. The inset of Fig. \ref{Fig3b} shows the Reynolds number $Re$  (i. e. descent velocity) to be decreasing with $l$. It is more than a factor of two between $Re$ of the diaspore with its natural stalk length and the diaspore with the shortest stalk length, as the fall velocity decreases when the diaspore rotates. 

In order to understand of what sets the minimal length of the stalk to reach steady autorotation, these experiments were reproduced with five different collected diaspores of common linden, and synthetic diaspores with the same camber, curvature and width of the wing, where only the length of the wing and the total weight of the diaspores were changed. For all these samples, we observe that $L_m$ decreases with $l$ until reaching a plateau and diverges for small values of $l$ (see the inset of Fig. \ref{Fig3c} for $L_m$ as a function of $l$ for all collected and synthetic diaspores). A natural length scale appears in the system, $L_N=U_s^2/g$, which describes the length at which a falling object will reach the vertical speed $U_s$, where $U_s$ is the steady descent velocity of the same diaspore with the longest stalk tested in our experiment. The torque applied on the wing by the aerodynamic forces is proportional to the position of the center of the wing $r$, see Fig \ref{fig2}b. We represent then the normalized transition length in Fig. \ref{Fig3c}, $L_m/L_N$ as a function of $l/r$ for all the experiments. We observe that the data groups around one curve where $L_m/L_N$ is close to 1 for long stalk and diverges when $l/r>1$. For steady autorotation to occur, the length of the stalk should be larger than about half of the span of the wing. If we return to our collected biological samples of \textit{Tilia x euopaea} diaspores in Fig. \ref{pdf}, we point out that this turns out to be the case for more than 99\% of the 859 samples.

\section{Conclusion}
We have shown that the vertical position of the center of mass and thus the length of the stalk is crucial for linden diaspores to rapidly induce a steady autorotating state. Decreasing the length of the stalk increases the decent distance before onset of autorotation, which can extend up to lengths comparable to the height of linden trees. Autorotation is crucial because it reduces the descent velocity of the diaspore by a factor two and as such increases the flight time.


The flow around model linden diaspores show a leading edge vortex and a trailing edge vortex around the rotating wing, which elevate the lift force on the diaspore. Because the wing of linden diaspore is curved, the leading edge vortex and the trailing edge vortex are concentrated towards the tip of the wing.  

The geometry of the collected samples from \textit{Tilia x europaea} indicates that linden diaspores grow sufficiently long peduncles to increase their chances for stable autorotating flight. Thus, the stalk length can be an evolutionary trait evolved to enable efficient flight of diaspores from linden trees.
Studies of the evolution of linden diaspores \cite{manchester1994inflorescence,Jia2021} show that laminar bracts, i.e., bracts extending above the intersection point with the peduncle $l_\text{wing}>0$, have appeared later and become more prevalent than non-laminar ones throughout the course of evolution. For a linden diaspore, having a laminar bract limits the torque from the lift force on the wing as it reduces $r$. The length of the stalk needed to guarantee quick onset of autorotation is then also decreased. The bract of linden diaspores may have evolved towards aerodynamically favourable properties through autorotating flight.

From an engineering view point, these passive flight designs have inspired bio-mimetic drone designs \cite{Kim2021,kim2024functional,Ulrich2010,Makdah_2019} used for a wide range of applications. Our work may inspire new avenues for design of bio-inspired flying devices and micro air vehicles, where changing the vertical position of the center of mass could be used to actively control the flight regime.

\begin{acknowledgments}
The research has been supported by the Research Council of Norway through its projects 301138 (NANO2021 program). We thank Olav Gundersen, Dr. Atle Jensen, Dr. Sif Fink Arnbjerg-Nielsen and Karen Samseth for their help with the experimental set-up and Dr. Stephane Poulain, Dr. Mathijs A. Janssen and Dr. Kaare H. Jensen for valuable discussions and input on our work.
\end{acknowledgments}

\appendix

\section{Methods}
\subsection{Data availability}All the data used in this paper is available on the platform sigma2 (DOI: https://doi.org/10.11582/2025.dy02we3f) \cite{Data}.
\subsection{Collection of biological samples}
\begin{figure}
    \centering
    \includegraphics[width=0.99\textwidth]{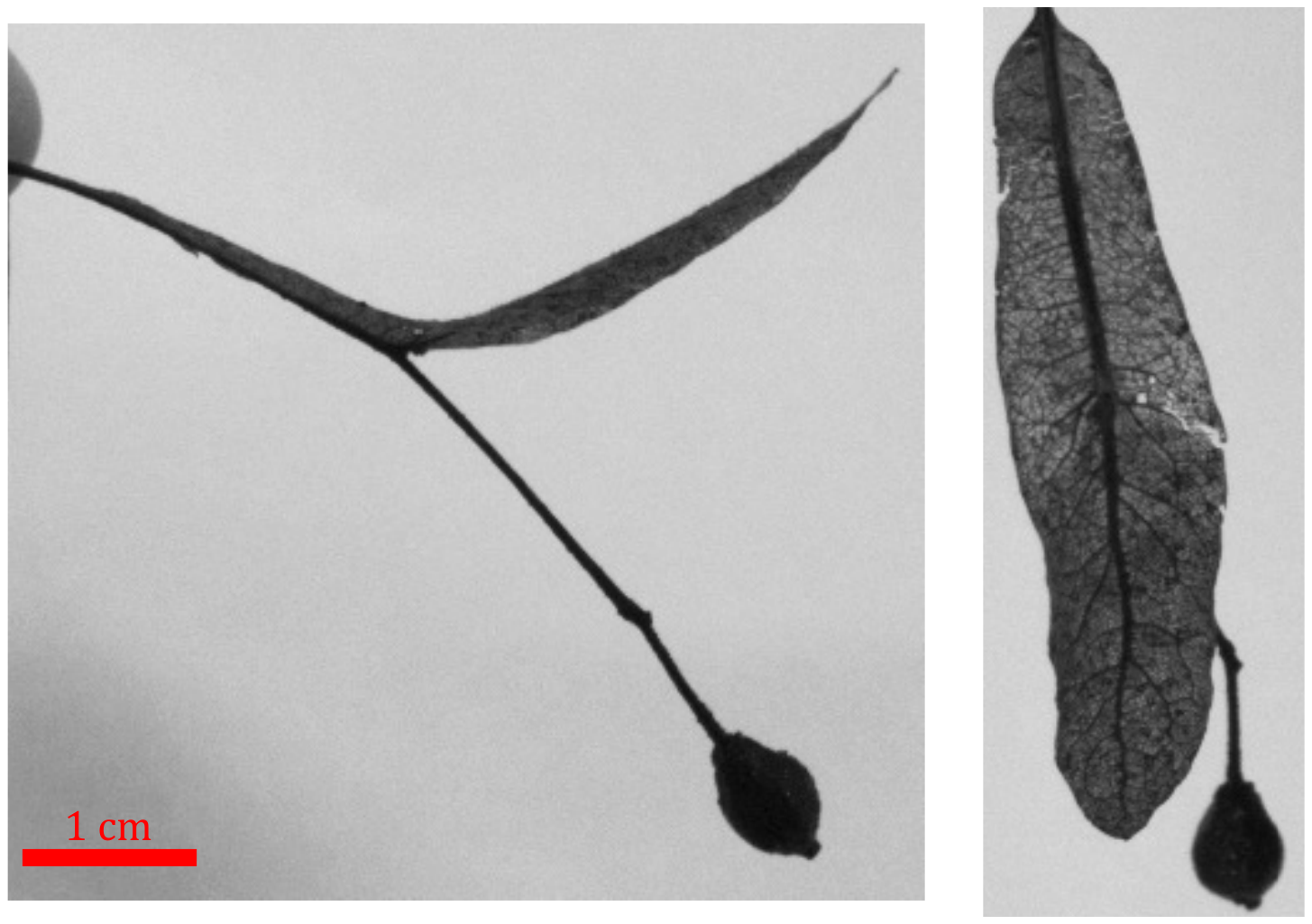}
    \caption{ Example of pictures from two different angles taken from the collected linden seeds to measure their geometrical properties.}
    \label{RealSeeds}
\end{figure}

A total of 859 biological samples of dispersal from the species \textit{Tilia x europeae} were picked up on the ground next to linden trees at five different locations in Oslo specified in Table \ref{tabel}. All found seeds were collected as samples as long as their stem and their wing were undamaged. We then took two images at different angles of the seeds to capture their geometrical properties (one example of such pictures is shown in Fig. \ref{RealSeeds}). 

\begin{table}[h!]
    \centering
    \begin{subtable}{0.99\textwidth}
        \centering
            \begin{tabular}{|c|c|c|}\hline
        Place & Number of diaspores collected & Geographical position \\\hline
        Kirkeveien & 321 & 59°56'09.6"N 10°43'54.6"E\\\hline
        Blindern Campus & 315 & 59°56'23.9"N 10°43'23.6"E\\\hline
        Jutulveien & 148 & 59°56'30.9"N 10°44'35.5"E\\\hline
        Sagene & 62 & 59°56'15.5"N 10°45'01.2"E\\\hline
         Ullevaal Sykehus & 13 & 59°56'07.5"N 10°44'36.7"E\\\hline     
    \end{tabular}
    \caption{Geographical locations where the biological samples where collected in Oslo.}
    \label{tabel}
        \end{subtable}
\begin{subtable}{0.99\textwidth}
    \centering
     \begin{tabular}{|c|c|c|c|c|c|}
  \hline
     &  $R_S$ (cm) & U (m/s) & $\Omega$(1/s) & Re & St  \\\hline
  Biological sample   & 2.5 - 6.8 & 0.5 - 3 &  60 - 190 & 3500 - 6000&0.5 - 0.8\\\hline 
     Synthetic diaspores in air &2 - 3.5 &1.5 - 2.5 & 140 - 250 &1500 - 4000 &0.5 - 0.75\\\hline
      Synthetic diaspores in water &2.5&0.16& 26 &4000 & 0.65\\\hline
   \end{tabular}
    \caption{Main characteristics of the biological samples and the synthetic diaspores used for the experiments in air and in water.}
   \label{Scaling}
\end{subtable}
\begin{subtable}{0.99\textwidth}
    \centering
     \begin{tabular}{|c|c|c|c|c|c|c|c|}
  \hline
   Symbol  &  Type of sample & Maximum stem length (cm) & $R_S$ (cm) & $r$ (cm) & $l_\text{wing}$ & $L_\text{wing}$\\\hline
  $\bigtriangledown$   & Biological & 6.4 & 6.8 & 3.5&2&6.5\\\hline 
  $\bigtriangleup$ & Biological & 3.8 & 6.4 &1.5&3.5 &4.5\\\hline
  $\vartriangleright$ & Biological & 4 & 5.1 &1.7&2.5 &3.8\\\hline
  $\vartriangleleft$ & Biological & 3.75 & 2.5 &1.45&3.8 &4.3\\\hline
  $\davidsstar$ & Biological & 3.5 & 5.4 &1&2.5 &4.2\\\hline  
  $\bigstar$ & Synthetic & 6 & 4.2 &&2 &4\\\hline  
    $\diamond$ & Synthetic & 4 & 1.6 &1&0 &2\\\hline  
  $\square$ & Synthetic & 6 & 4.3 &1.75&2 &5\\\hline  
    $\bigcirc$ & Synthetic & 6 & 3.5 &1.5&1 &4\\\hline  
   \end{tabular}
    \caption{Geometrical properties of the samples used in the experiments with their symbols in Fig 3b.}
   \label{Scaling}
\end{subtable}
\label{tables}
\caption{}
\end{table}

\subsection{Flight experiments with collected samples and synthetic diaspores}

Four of the collected biological samples were used to perform flight experiments. We also used synthetic diaspores with controlled geometrical properties for flight experiments. In order to make these synthetic diaspores, we print a cambered and curved rectangular mold with a Form 3 3D-printer by Formlabs similar to the biological samples. We then use the mold to make a papier-maché cambered and curved wing made of two layers of tissue paper which we then coat with transparent lacquer. The stalk is 3D-printed and glued to the wing. The weight is made of Blu-Tack and attached to the end of the stalk. 

To evaluate the role of the position of the center of mass for both biological samples and synthetic diaspores, we reduce gradually the length of the stalk by cutting its end. We keep the weight constant by adding Blu-Tack to compensate for the weight of the cut part of the stalk. The diaspores thus prepared are released from a clamp placed near the ceiling of the laboratory and fall in still air until they reach the floor 2 meters down. To test the relevance of our results over scales closer to the height of a linden tree, we also performed flight experiments over the mezzanine of our laboratory (up to 5 meters) and in the stairs of the building (up to 9 meters).

\begin{figure}
\centering
\begin{subfigure}{0.24\textwidth}
\includegraphics[width=\textwidth]{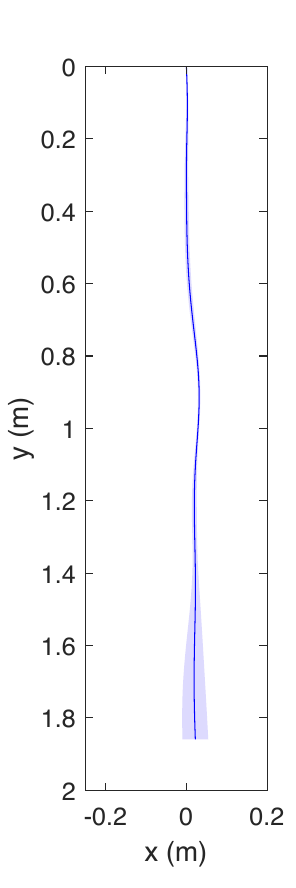}
\caption{}
\end{subfigure}
\hfill
\begin{subfigure}{0.24\textwidth}
\includegraphics[width=\textwidth]{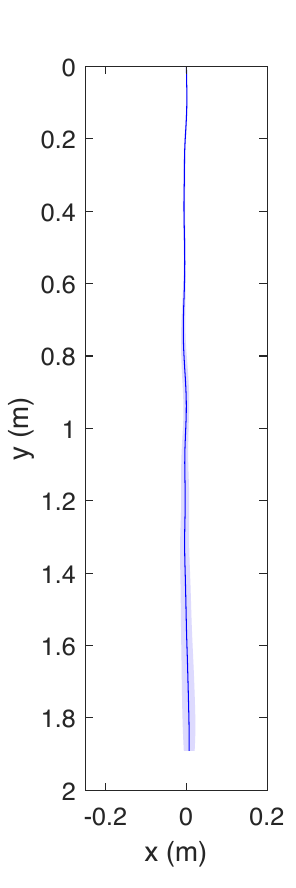}
\caption{}
\end{subfigure}
\hfill
\begin{subfigure}{0.24\textwidth}
\includegraphics[width=\textwidth]{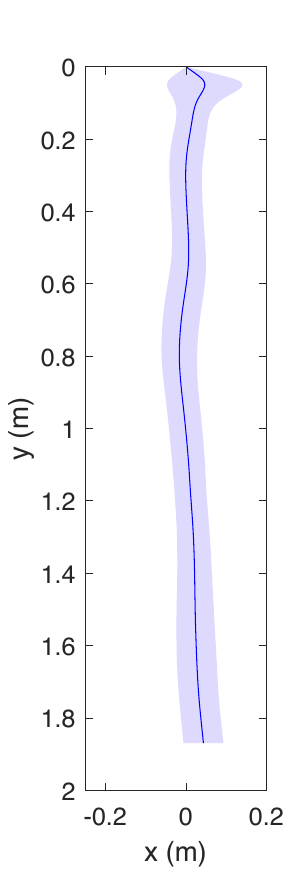}
\caption{}
\end{subfigure}
\hfill
\begin{subfigure}{0.24\textwidth}
\includegraphics[width=\textwidth]{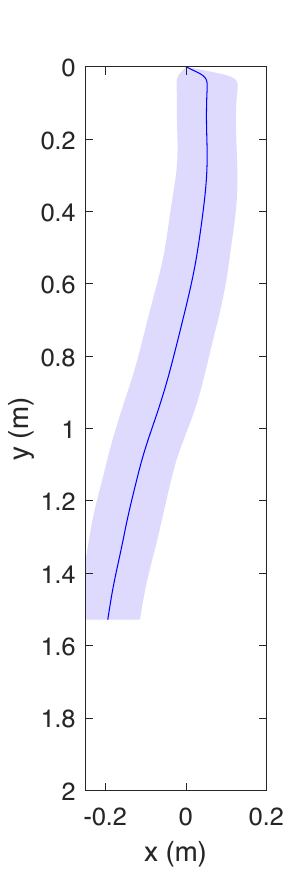}
\caption{}
\end{subfigure}
\caption{Trajectories of the pods for different stalk length. (a) $L=6$ cm. (b) $L=3$ cm.  (c) $L=2.5$ cm. (d) $L=2$ cm. The line represents the mean trajectory of the 5 repetitions of the experiment and the blue area represents the standard deviation over those repetitions. For longer stalks ((a) and (b)), there is very little variation between the repetitions of the experiment and the path is almost vertical. For longer stalks (c) and (d), the variations between the repetitions of the same experiment increase and the diaspore glide more in the horizontal direction.}
\label{Traj}
\end{figure}

The movement of the diaspores is recorded with a high speed camera (Photron Fastcam mini) at 500 fps through a 28 mm lens. The images of the fall are binarized and we track the position of highest and lowest point of the diaspore. Examples trajectories of the pods for a synthetic diaspore with different stalk lengths are shown in Fig. \ref{Traj}. We observe that for longer stalks ((a) and (b)), the diaspore follows almost identical path for each repetition of the experiment and that the trajectory deviates only slightly from the vertical. For longer stalks, the path of the diaspore becomes more sensitive to small changes in the initial condition, there is therefore more deviation between the repetitions of the experiments. The horizontal variations of the path become also more significant for shorter stalks, the trajectory is not purely vertical.

\begin{figure}
\centering
\begin{subfigure}{0.48\textwidth}
\includegraphics[width=\textwidth]{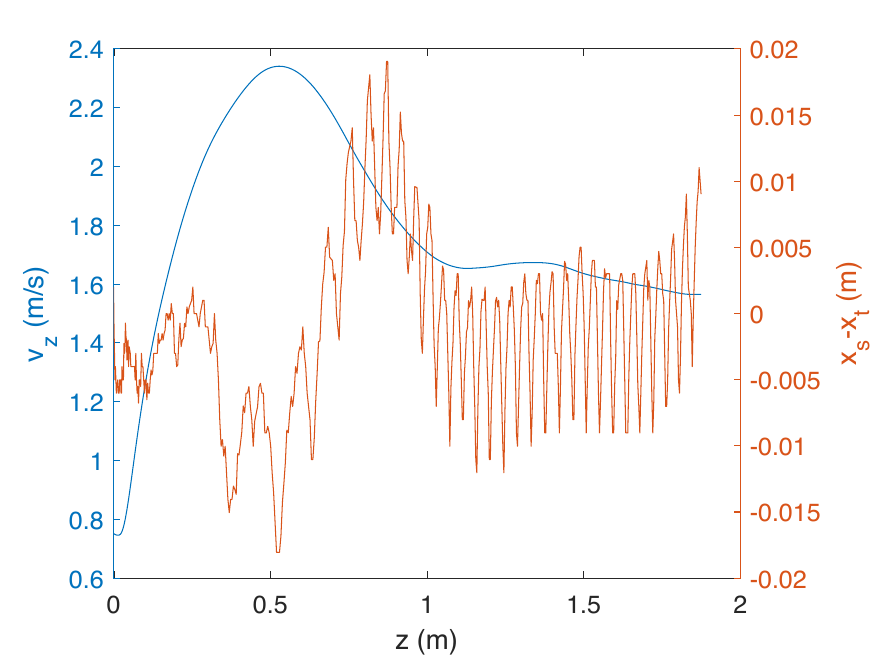}
\caption{}
\end{subfigure}
\hfill
\begin{subfigure}{0.48\textwidth}
\includegraphics[width=\textwidth]{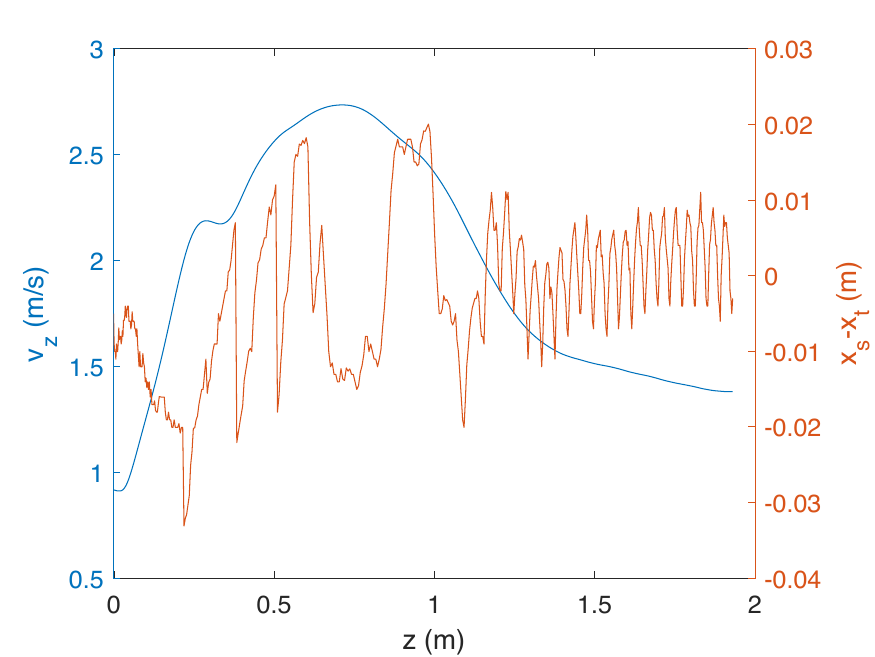}
\caption{}
\end{subfigure}
\hfill
\begin{subfigure}{0.48\textwidth}
\includegraphics[width=\textwidth]{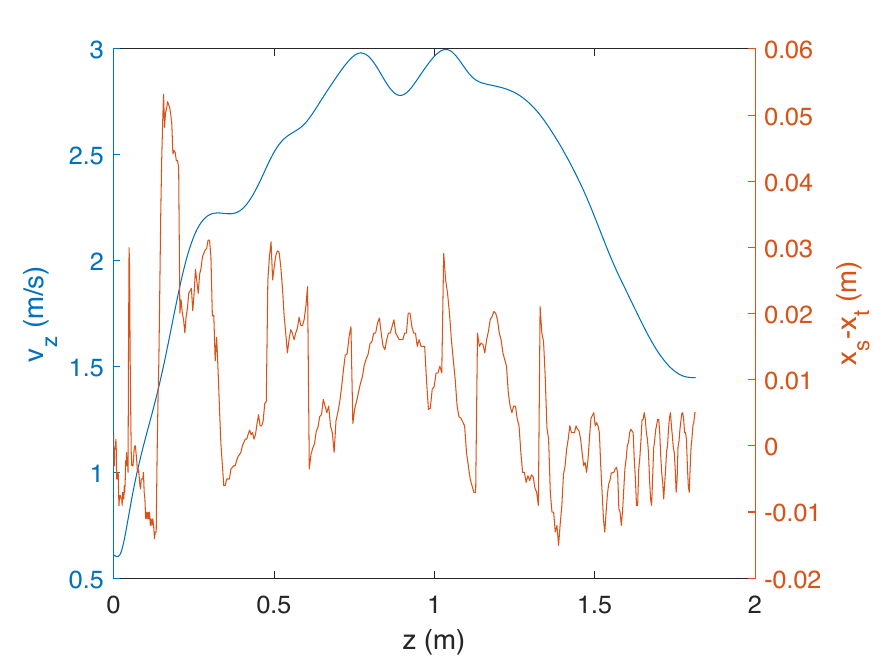}
\caption{}
\end{subfigure}
\hfill
\begin{subfigure}{0.48\textwidth}
\includegraphics[width=\textwidth]{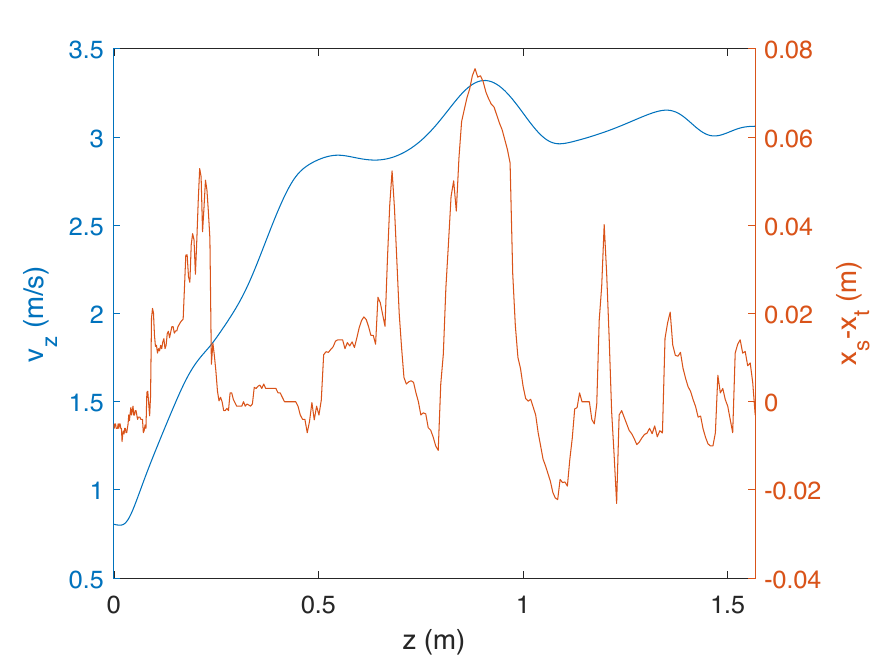}
\caption{}
\end{subfigure}
\caption{Blue: Vertical velocity of the pods as a function of the vertical position, orange: relative horizontal positions of the tip and the pod. (a) $L=6$ cm. (b) $L=3$ cm.  (c) $L=2.5$ cm. (d) $L=2$ cm. For short stalks, in (a), the diaspore accelerates when released, then the velocity reaches a maximum. When the diaspore starts spinning, we see regular oscillations of the curve of the horizontal position. We observe that the autorotation starts when the diaspore decelerates. For longer stalks, the autorotation and the deceleration starts after a longer fall (in (b) and (c), and eventually is not seen over the height of the experiment (d).}
\label{vz}
\end{figure}
\subsection{PTV measurements around synthetic fruits in water}

\begin{figure}
    \begin{subfigure}{0.82\textwidth}
\includegraphics[width=0.99\textwidth]{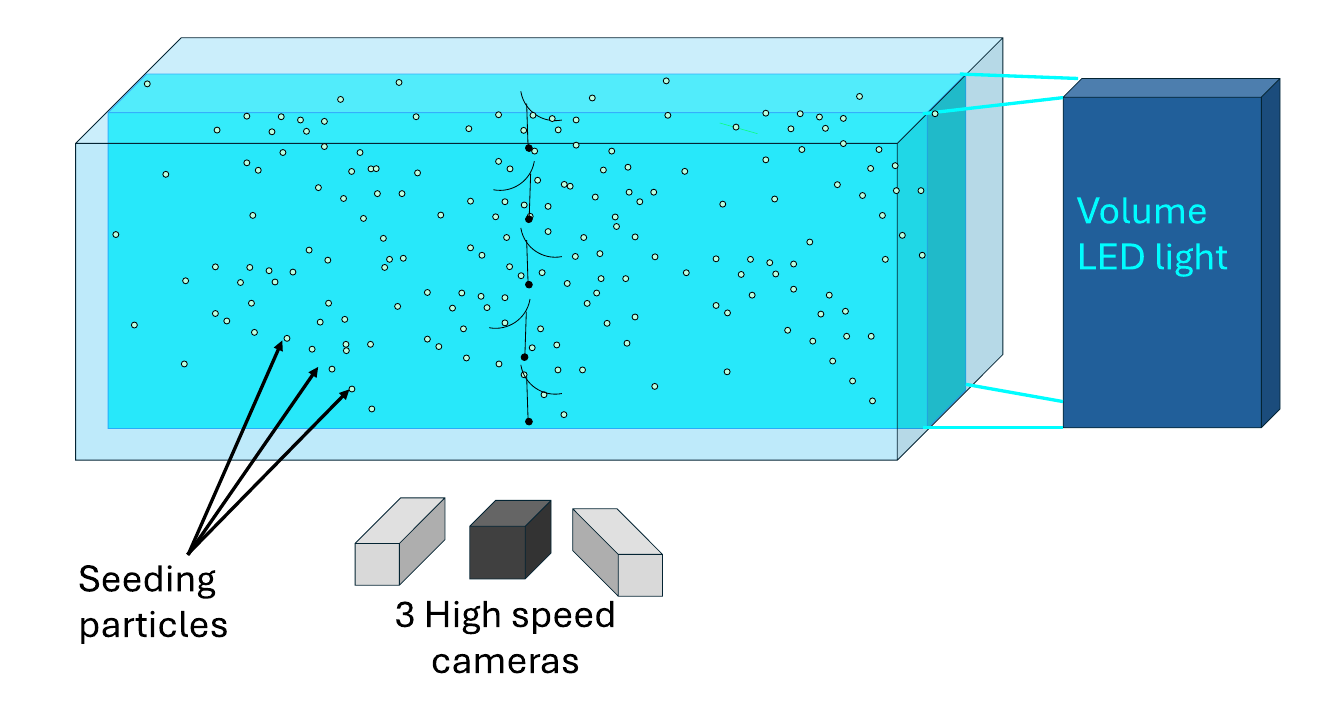}
\caption{}
\label{PTVsetup}
\end{subfigure}
    \begin{subfigure}{0.16\textwidth}
\includegraphics[width=0.99\textwidth]{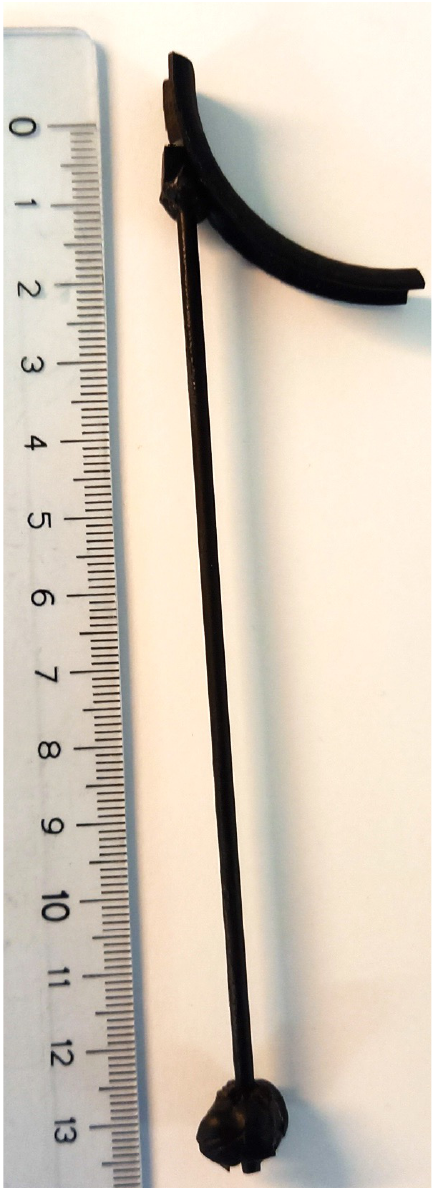}
\caption{}
\label{PTVseed}
\end{subfigure}
\caption{(a) Set-up for the 3D flow measurements in water. (b) Synthetic diaspore used for the measurements in water.}
\label{PTV}
\end{figure}
In order to observe the structure of the fluid flow around the wing, we performed flight experiments in a water tank with synthetic diaspores.
The curved and cambered wing of the diaspore was printed with our Form3 3D-printer by Formlabs. A stainless steel stalk of diameter of 3 mm was then attached to it. Finally, a fishing weight was glued to the stalk to model the weight of the pods. An image of the synthetic diaspore can be found in Fig. \ref{PTVseed}. 
The fluid was seeded with polyamid tracer particles from Dantec of diameter 60 $\mu$m. A volume of the fluid was illuminated by an array 72 high-power LEDs in an area of 10 cm x 30 cm (LED-Flashlight 300 by La Vision).
The movement of the particles in the lighted volume was captured by three high speed cameras at a frame rate of 200 fps (two Photron Nova R3 on the sides and one Photron Fastcam mini in the middle, see Fig \ref{PTVsetup} for the set-up). 
The images were analyzed by the Shake The Box algorithm of La Vision to get the positions of the particles in each frame. The vector field was then reconstructed from the particles' tracks.

\subsection{Scaling analysis}

The main characteristics of the biological samples and the synthetic diaspores in air and water are shown in table \ref{Scaling}. The dimensions and weight of the synthetic diaspores were chosen so that the Reynolds and the Strouhal numbers would be of the same order as the one of the biological samples. As water is more viscous than air, the synthetic diaspore in water has to fall slower than the biological samples in air so that the Reynolds number are of the same order of magnitude. The rotation speed $\Omega$ has then to be reduced by the same order for the Strouhal numbers to also match. 
\subsection{Steady state}

\begin{figure}
\centering
\begin{subfigure}{0.49\textwidth}
\includegraphics[width=\textwidth]{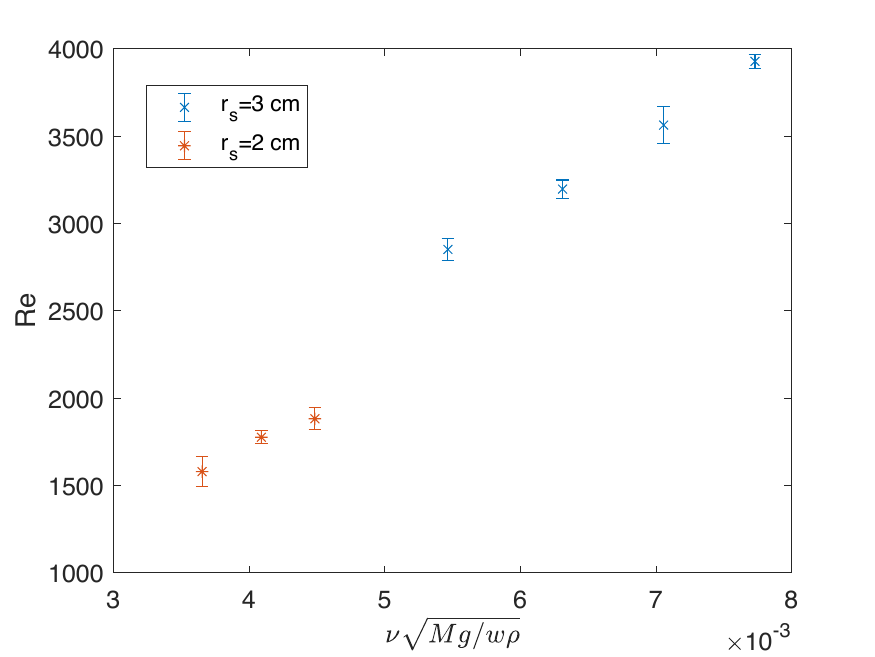}
\caption{}
\label{re}
\end{subfigure}
\hfill
\begin{subfigure}{0.49\textwidth}
\includegraphics[width=\textwidth]{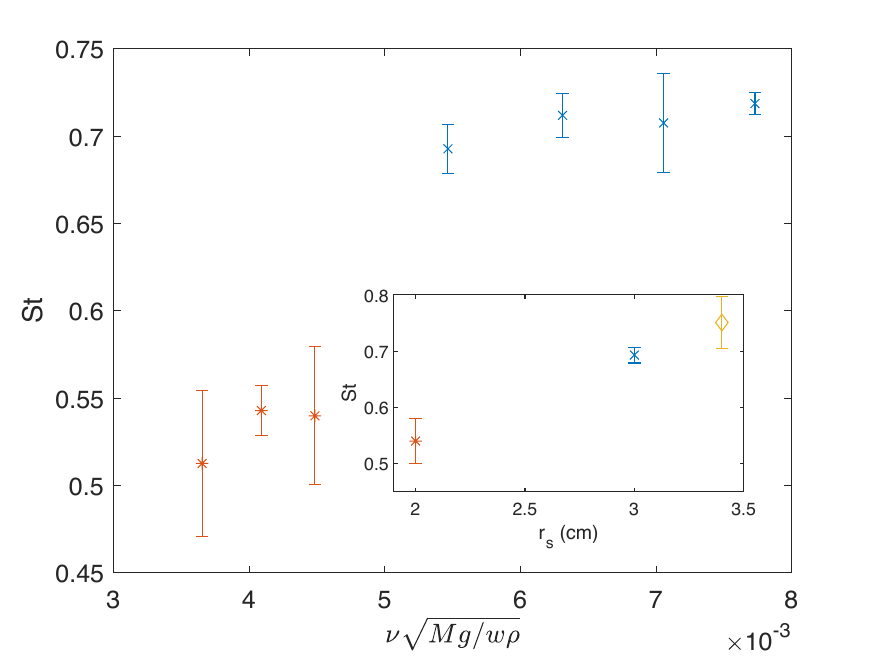}
\caption{}
\label{st}
\end{subfigure}
\caption{(a) $Re$ as a function of $\nu\sqrt{M g/w\rho}$ for two different geometries. (b)  $St$ as a function of $\nu\sqrt{M g/w\rho}$ for two different geometries, insert: $St$ as a function of $r_S$.}
\label{ReSt}

\end{figure}
Once they have reached their steady state the diaspores fall at a constant velocity $U$ and rotate at a constant angular velocity $\Omega$. From $U$ and $\Omega$, we get $Re$ and $St$ for diaspores of different weight and different geometry. If we balance the vertical aerodynamic force with the weight, the vertical velocity should follow the simple relationship \cite{Norberg1973}:
\begin{equation}
U\propto\sqrt{mg/\rho A},
\end{equation}
then,
\begin{equation}
Re\propto \nu\sqrt{Mg/\rho w}.
\label{steady}
\end{equation}

We varied the weight for two different wing length with synthetic diaspores. As shown in Fig \ref{re}, for the two geometries tested, the experimental data follows the equation \ref{steady}. The Strouhal number does not depend on the weight of the diapores and seems to increase with $R_S$ (see Figure \ref{st}).
\subsection{Blade Element Theory}
\begin{figure}
\centering
\begin{subfigure}{0.4\textwidth}
  \includegraphics[width=1\textwidth]{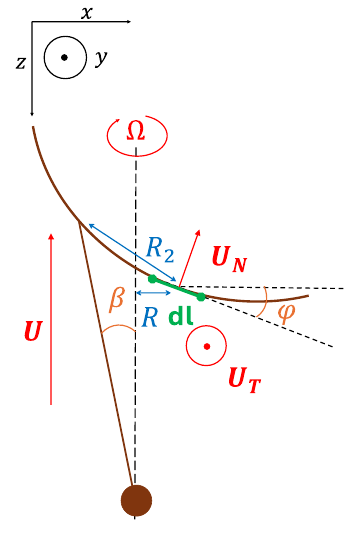}  
  \caption{}
  \label{LBTd}
\end{subfigure}
\begin{subfigure}{0.4\textwidth}
  \includegraphics[width=1\textwidth]{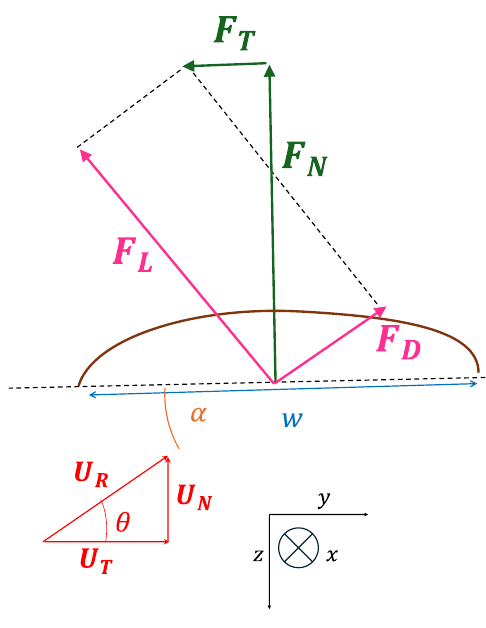} 
  \caption{}
  \label{LBTd2}

\end{subfigure}
  \caption{(a) and (b) Schematic of the diaspore with its main geometric properties and representation of the aerodynamic forces acting on the wing from two different view points. }
\end{figure}
\begin{figure}
\begin{subfigure}{0.49\textwidth}
  \includegraphics[width=1\textwidth]{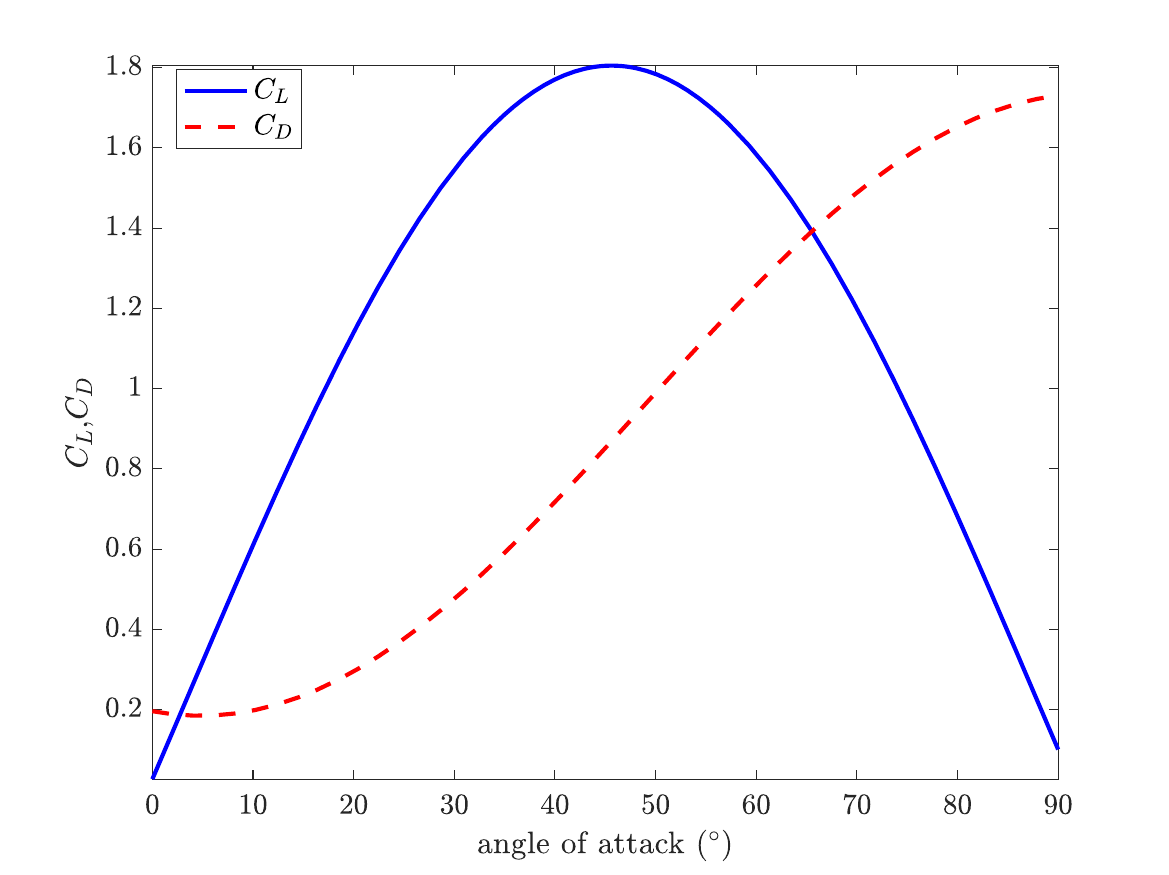}  
    \caption{}
\label{LBTc}
\end{subfigure}
\begin{subfigure}{0.49\textwidth}
  \includegraphics[width=1\textwidth]{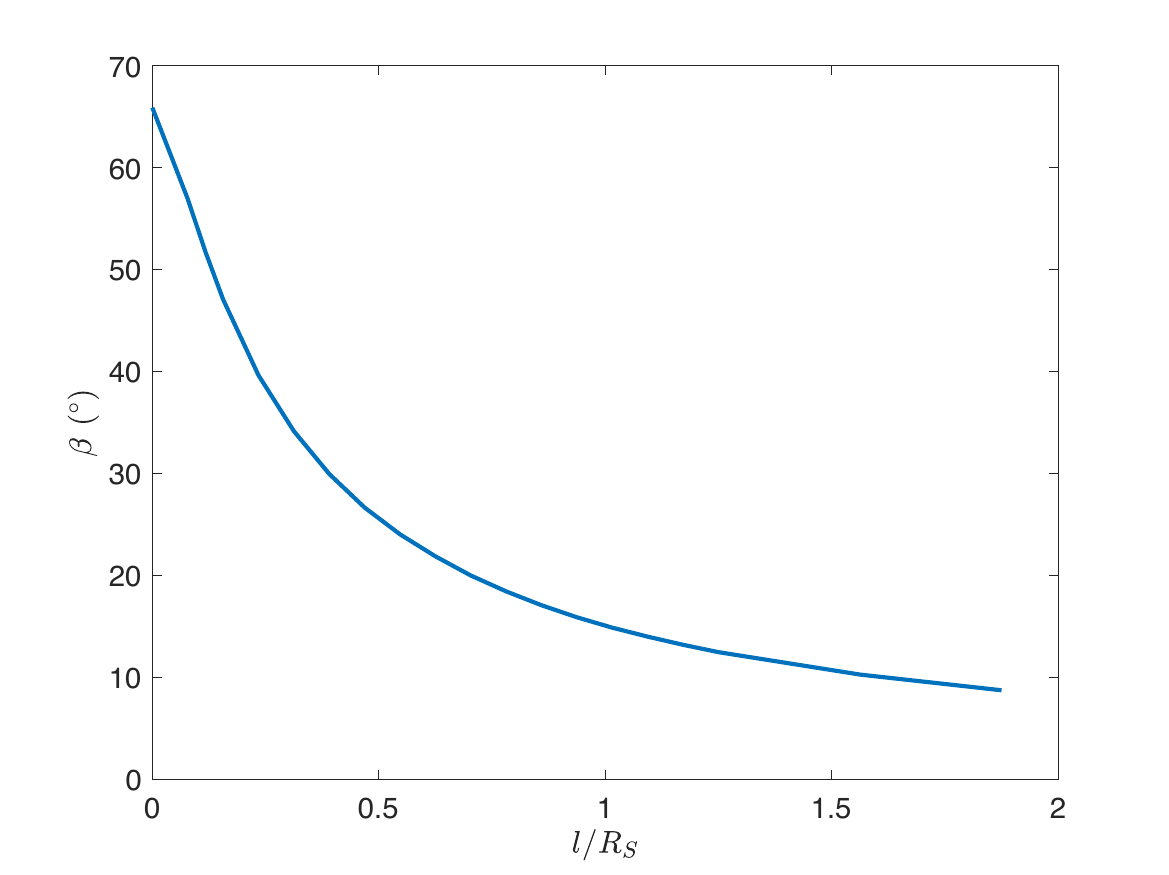}  
    \caption{}
\label{LBTb}
\end{subfigure}
\begin{subfigure}{0.49\textwidth}
  \includegraphics[width=1\textwidth]{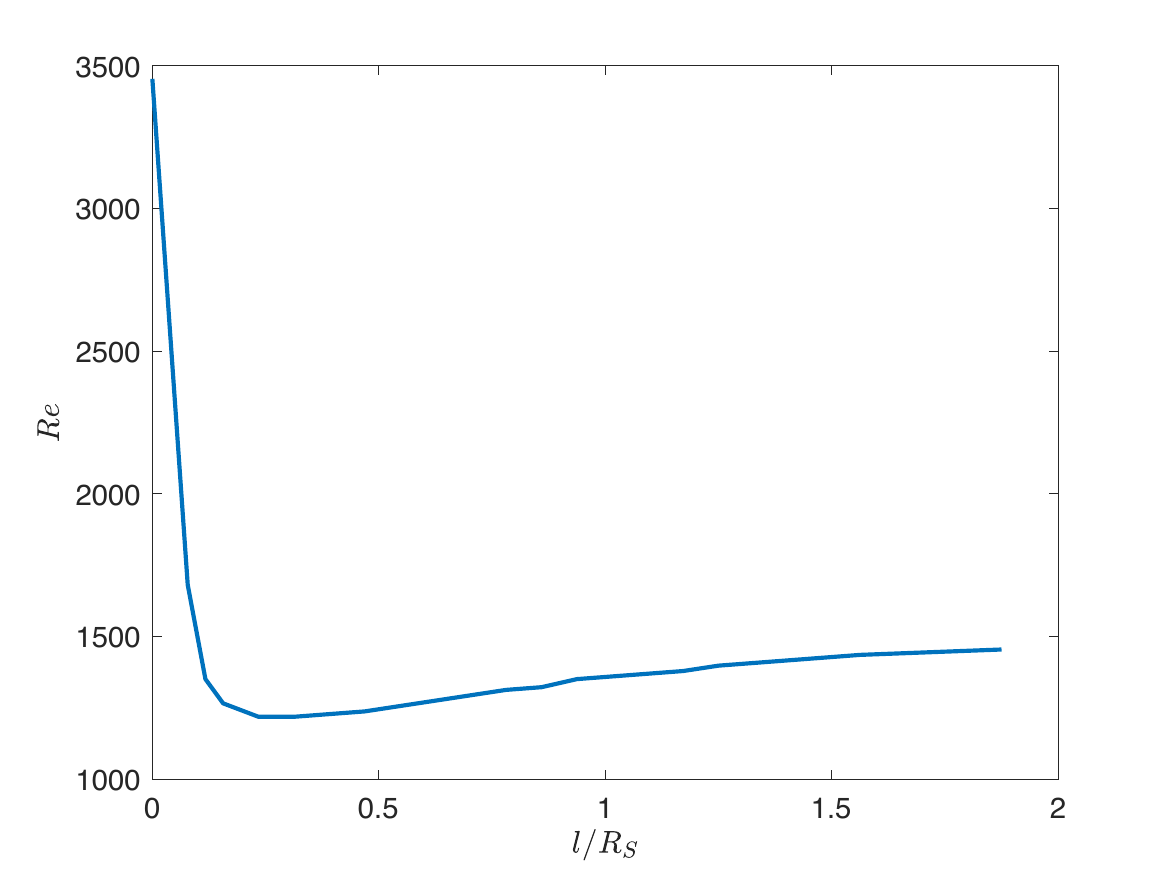} 
    \caption{}
  \label{LBTRe}
\end{subfigure}
\begin{subfigure}{0.49\textwidth}
  \includegraphics[width=1\textwidth]{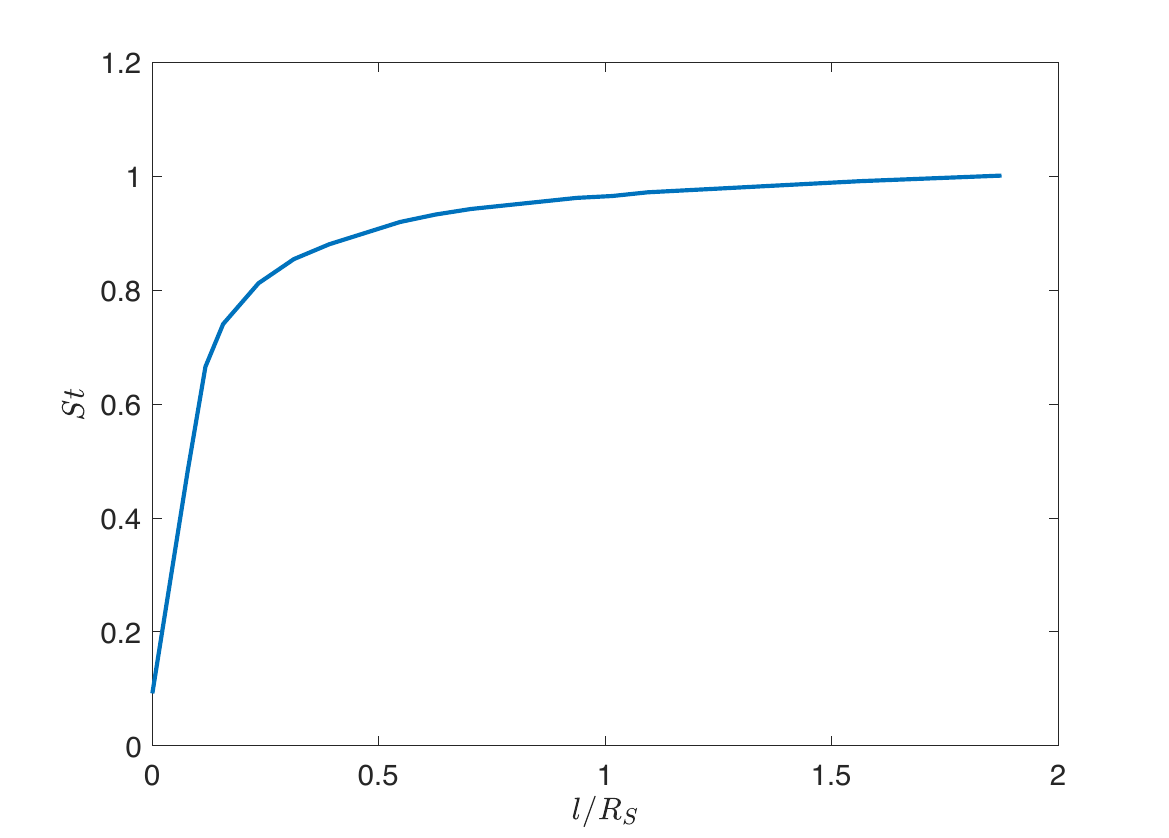}  
    \caption{}
\label{LBTSt}
\end{subfigure}

\caption{(a) Values of the lift and drag coefficients as a function (b), (c) and (d) value of $\beta$, $Re$ and $St$ at the equilibrium points as a function of the length of the stalk.}
\label{BET}
\end{figure}
To describe the steady state motion of the linden diaspore, we compute the contribution of the aerodynamic forces by using the blade element theory. We decompose the wing in thin element of length $dl$. The draft and the lift forces, $F_D$ and $F_L$ are calculated locally on each blade element and then their contribution is integrated over the length of the wing to get the different components of the forces and momenta.

\begin{equation}
F_L=\frac 1 2 \rho U_r^2  w dl C_L(\alpha),
\label{F_L}
\end{equation}

\begin{equation}
F_D=\frac 1 2 \rho U_r^2  wdl C_L(\alpha),
\label{F_D}
\end{equation}

where $\rho$ is the density of the fluid, $U_r$ is the local relative wind velocity, $wdl$ is the surface area of the blade element, $\alpha$ the local of attack and $C_L(\alpha)$, $C_D(\alpha)$ are the local lift and drag coefficients. $C_L$ and $C_D$ are computed as suggested by \cite{Fauli2019} using equation 12 and 13 of \cite{Wang2004} (see Fig. \ref{LBTc}). We also correct the lift coefficient to take into account camber, by adding on offset attack angle $\Delta \alpha$ as suggested in \cite{Fauli2019}.

We place ourself in the frame of reference of the falling and rotating diaspore. We assume that the diaspore is falling at a constant velocity $U$ and rotating, at a constant speed $\Omega$ around a vertical axis positioned at the center of mass of the diaspore. The stalk has a constant tilt angle $\beta$ around the $y$ axis. We assume that there is no tilt around the $x$ axis and that the axis of rotation is included in the plane of the center of the diaspore, as it is the case experimentally.  We calculate the vertical force $F_z$, the momentum along the axis of rotation, $M_z$, and the momentum along the tilt axis $M_y$ to obtain the equilibrium points. 
Those can be expressed as a function of the tangential and normal components of the aerodynamic forces, $F_T=F_L \sin \theta-F_D \cos \theta$ and $F_N=F_L \cos \theta+F_D \sin \theta$ (see Figure \ref{LBTd}). 

\begin{equation}
F_z=-\int_{-l_\text{wing}}^{L_\text{wing}} F_N \cos \varphi dl +mg 
\label{Fz}
\end{equation}

\begin{equation}
M_z=\int_{-l_\text{wing}}^{L_\text{wing}} F_T R dl
\label{Fz}
\end{equation}

\begin{equation}
M_y=\int_{-l_\text{wing}}^{L_\text{wing}} F_T R_2 \sin \psi dl+mg \sin \beta L_\text{stalk}-m L_\text{stalk}^2 \sin^2\beta \Omega^2 \cos \beta
\label{Fz}
\end{equation}

 We can then find the equilibrium values $U_\text{eq}$, $\Omega_\text{eq}$ and $\beta_\text{eq}$ for which $F_z$, $M_z$ and $M_y$ reach zero simultaneously. 

 A scan of the parameters reported in the main text is made with the theoretical model with the simplification of the geometry, which consists only of a rectangular curved and cambered geometry. Fig. \ref{LBTb}, \ref{LBTRe}, \ref{LBTSt} shows the results for different stalk lengths $l$, where as the experiments we notice the descent velocity increases as the stalk shortens and it plateaus when $l/R_s>1$. As explained qualitatively in the main article, the angle necessary to get the momentum around the horizontal axis to be balanced decreases when $l$ increases (see Fig. \ref{LBTb}). The rotation speed and thus the Strouhal number increases with the length of the stalk, when the Reynolds number reaches a minimum value and then increases very slowly with the length of the stalk. 

%

\end{document}